\documentclass{article}

\usepackage[english]{babel}

\usepackage[letterpaper,top=2cm,bottom=2cm,left=3cm,right=3cm,marginparwidth=1.75cm]{geometry}

\usepackage{amsmath}
\usepackage{amsfonts}
\usepackage{amssymb}
\usepackage{graphicx}\usepackage{stackengine}
\usepackage[colorlinks=true, allcolors=blue]{hyperref}
\usepackage{natbib}

\usepackage[dvipsnames]{xcolor}
\title{What the flock knows that the birds do not: exploring the emergence of joint agency in multi-agent active inference}

\author{Domenico Maisto, Davide Nuzzi, Giovanni Pezzulo$^{*}$  \\
\\
Institute of Cognitive Sciences and Technologies, National Research Council, Rome, Italy  \\
$*$ Corresponding author: giovanni.pezzulo@istc.cnr.it}

\date{}

\begin{document}
\maketitle

\begin{abstract}

Collective behavior pervades biological systems, from flocks of birds to neural assemblies and human societies. Yet, how such collectives acquire functional properties—such as joint agency or knowledge—that transcend those of their individual components remains an open question. Here, we combine active inference and information-theoretic analyses to explore how a minimal system of interacting agents can give rise to joint agency and collective knowledge. We model flocking dynamics using multiple active inference agents, each minimizing its own free energy while coupling reciprocally with its neighbors. We show that as agents self-organize, their interactions define higher-order statistical boundaries (Markov blankets) enclosing a “flock” that can be treated as an emergent agent with its own sensory, active, and internal states. When exposed to external perturbations (a “predator”), the flock exhibits faster, coordinated responses than individual agents, reflecting collective sensitivity to environmental change. Crucially, analyses of synergistic information reveal that the flock encodes information about the predator’s location that is not accessible to every individual bird, demonstrating implicit collective knowledge. Together, these results show how informational coupling among active inference agents can generate new levels of autonomy and inference, providing a framework for understanding the emergence of (implicit) collective knowledge and joint agency.

\end{abstract}


\section{Introduction}

\begin{quotation}

The anchor of all my dreams is the collective wisdom of mankind as a whole. -- \emph{Nelson Mandela}
\end{quotation}

The fundamental unit of analysis in biology and cognition is often the \emph{agent}---an entity, such as a person, animal, or even a single neuron or cell, that possesses a well-defined boundary and some degree of \emph{agency}, that is, autonomy in perceiving and interacting with an environment. However, many of the most intriguing phenomena in biological and cognitive systems emerge not from isolated agents but from the collective dynamics of multiple interacting ones. Examples include the coordinated behaviors and self-organization of cells composing a body morphology \citep{levin2019computational,friston2015knowing,pezzulo2016top,levin2017endogenous,manicka2025field}neural population dynamics underlying brain function \citep{freeman1994characterization,fuchs1992phase,friston2012perception,deco2015rethinking,hesse2014self}, the collective behaviors animals \citep{holldobler2009superorganism,krause2002living,sumpter2010collective}, distributed models of mind \citep{minsky1986society,hofstadter1999godel}, and even human societies viewed as collective agents \citep{sawyer2005social}. Such \emph{collective intelligence} phenomena occur across multiple scales of biological and cognitive organization \citep{mcmillen2024collective}.
 
These examples suggest that the concept of agency can be generalized beyond single individuals or parts, extending to systems composed of many interacting components that together behave \emph{as if} they were one agent. If agency is defined as the capacity to sense, infer, and act purposefully, then \emph{joint agency} (also called \emph{shared} or \emph{collective agency}) arises when perception, cognition, and action extend beyond a single entity---emerging instead from the coordinated dynamics of multiple decision-making units.

Notions of joint agency appear at many levels of organization and cognitive sophistication. At the level of human social behavior, classical theories of social cognition emphasize both social interaction and collective behavior: while individuals act autonomously, they also participate in shared social structures. Theories of \emph{mentalizing} highlight that agents form beliefs about, and models of, other agents' minds \citep{frith2005theory}. Complementary accounts emphasize \emph{shared and aligned representations}, in which multiple agents maintain overlapping internal models of their environment \citep{pezzulo2025predictive}. In these frameworks, joint agency corresponds to phenomena such as shared representations \citep{sebanz2006joint,Sebanz2021}, collective intentions and goals \citep{Bratman2013,butterfill2023towards,tomasello2005understanding}, joint commitments \citep{gilbert2020social}, common ground \citep{Clark1996}, joint payoffs and team reasoning \citep{Sugden2003}, and ``we-representations'' of action and intention \citep{Gallotti2013}. Empirical work in cognitive neuroscience supports these views, identifying mechanisms such as \emph{neural coupling} \citep{Hasson2012,Hasson2016,Marsh2009,Keller2016}, \emph{mirroring} \citep{Gallese2004,rizzolatti2004mirror} and \emph{sensorimotor communication} \citep{Pezzulo2019} as enabling the fine-grained spatiotemporal coordination required for joint action. Subjectively, such coordination can give rise to a \emph{shared sense of agency}---the feeling that outcomes were caused \emph{together} \citep{pacherie2014does}.

Computational models have begun to formalize these social dynamics in terms of interacting agents that communicate, coordinate, and collaborate toward shared objectives. These frameworks typically assume internal (generative) models that encode distinctions between self and others---``my,'' ``your,'' and ``our'' actions and intentions \citep{Wolpert2003,kilner2007mirror}. More recent approaches, however, explore \emph{interactive} dynamics in which multiple agents maintain and update a \emph{shared world model} to minimize prediction errors toward a common goal \citep{Friston2015,Maisto2023,Friston2024}. Related concepts include the notion of \emph{agent-neutral} models, or internal models that predict the collective consequences of joint actions regardless of who executed them \citep{pezzulo2017avoiding,pezzulo2025predictive}, and \emph{shared beliefs} such as ``public beliefs'' \citep{Foerster2019} or the ``imagined we'' \citep{Tang2022}, where collective cognitive dimensions (e.g., beliefs, plans, agency)---shared across multiple individuals---supersede and drive individual cognitive states.


Here we are concerned with a more primitive notion of joint agency that arises at lower levels of organization: at the level of collective behavior and self-organization among simple, particle-like agents that lack advanced cognitive abilities and rich internal models incorporating notions like ``our beliefs'' or ``our plans''. This primitive form of joint agency emerges from the simple fact that teams of agents can infer their own states and actions based on those of other, surrounding agents to which they are informationally coupled---and they are capable of making collective decisions \citep{couzin2005effective,sridhar2021geometry}. There is a long tradition of studying the self-organization of collectives, such as active particles, animal swarms, and robot ensembles, using methods from statistical physics and information theory \citep{dorigo2021swarm,bialek2012statistical,cavagna2010scale,bechinger2016active,gomez2023fish,abdel2022self,durve2020learning}. Complementary approaches have also been developed to quantify \emph{causal emergence} and the extent to which higher-level collective dynamics exhibit causal power beyond that of their individual components \citep{hoel2013quantifying,rosas2020reconciling}.

An emerging trend is the study of collective phenomena and multi-agent systems within the active inference framework \citep{Parr2022,Friston2015,Maisto2023,heins2024collective,levchuk2019active,beckenbauer2025orchestrator,ruiz2024factorised}. Active inference was initially developed to address the cognitive and neural processes associated with isolated biological organisms and their action-perception cycle. The general idea was that a single imperative or objective function --- the minimization of variational free energy --- suffices to explain both perception and action and their associated neural dynamics in biological organisms. 

Recent developments have extended active inference principles to the collective dynamics of multiple agents, simpler (e.g., active particles) or more complex. The main difference between active inference and typical statistical approaches used to study collective phenomena is that each component is a full-fledged agent, with its action-perception cycle and minimizing its free energy based on local signals from the environment and/or other agents. For instance, in a network of agents playing the role of ``neurons,'' each minimizing its own variational free energy, a single neuron can infer whether or not to fire based on the activity of surrounding neurons. The ensemble of neurons can thereby exhibit synchronous dynamics and become collectively responsive to perceptual stimuli and reward contingencies \citep{palacios2019emergence,gandolfi2022emergence,gandolfi2025network}. Similarly, in a collection of cells engaged in morphogenesis and pattern formation, each cell can infer its own position in the final body morphology from the chemical signals emitted by neighboring cells, while simultaneously emitting signals that guide others \citep{friston2015knowing}. Functionally, each cell acts as an individual active inference agent minimizing its own free energy, yet its collective dynamics lead to the emergence of a coherent body morphology that is resilient---for example, capable of reconstructing itself after perturbation. 

When endowed with joint agency, active inference agents do not lose their individual autonomy: each continues to infer and minimize its own free energy. However, their collective agency at the system level supersedes individual agency to some extent, as the fate of each agent becomes jointly determined by the states and signals of others \citep{heins2024collective}. This formulation allows studying agency at two (or more) nested levels: the level of the single agent and the agent collective (and of a collective composed of collectives, and so on). The formulation also allows studying how collective processes can go awry, for example, when a single agent becomes insensitive to signals from other agents and hence myopically pursues individual rather than collective goals. In this perspective, a breakdown of cell-cell communication that causes individual cells to prioritize unicellular objectives rather than large-scale, collective morphogenetic goals has been proposed as a possible mechanism for cancer \citep{levin2021bioelectric,levin2021bioelectrical}.

Despite significant progress in modeling the collective behavior of multiple (simple) active inference agents, the relationships between these models and the broader notions of joint agency remain only partially understood. Previous active inference simulations have primarily focused on the self-organization of agents into cohesive multi-agent structures---such as bodies or coordinated ensembles---but have paid less attention to the functional consequences of this self-organization for higher-order phenomena such as \emph{joint agency} or \emph{collective beliefs}. These functional notions are typically not explicitly encoded in the internal generative models of the simulated agents, unlike in active inference models of higher-level cognition (e.g., human--human joint action \citep{Maisto2023}), where such constructs are explicitly represented. 

This raises an important question: can a collective system, even in the absence of explicit cognitive representations, be said to possess a form of knowledge or operational capability that extends beyond that of its individual components? In other words, can the ensemble as a whole instantiate a \emph{collective world model}---a functional integration of information and inference processes that confers emergent, joint agency?

To address these questions, we present and analyze a simple self-organizing multi-agent simulation inspired by flocking dynamics, in which each ``bird'' is modeled as an active inference agent minimizing its own variational free energy. In this framework, each bird updates its beliefs about hidden states (e.g., its heading direction) based on local observations and acts to minimize prediction error, leading to emergent coordination as birds gradually align their trajectories over time. 

We first illustrate how a formal notion of both individual agency (of the birds) and joint agency (of the flock) can be derived within this setting, using the concept of a \emph{Markov blanket} to delineate the statistical boundaries between agents, their interactions, and the environment. We then present simulations in which a ``predator'' is introduced to perturb and destabilize the flock, allowing us to examine two distinct phases: one in which the flock exhibits joint agency, and another in which it does not. This minimal model allows us to visualize the transition from individual to joint agency (in the absence of perturbations) and its dissolution under external disruption (when the predator attacks). It thereby demonstrates how birds can dynamically merge into, or separate from, a collective agent---a \emph{flock}---through changing patterns of coupling and inference. Finally, we employ \emph{synergistic information}---a quantitative measure of how information is distributed and integrated across multiple components---to assess the extent to which the collective flock possesses implicit ``knowledge about'' the predator’s position that exceeds the information explicitly represented in the internal models of individual birds. This approach enables us to characterize the emergence of system-level inference and coordination, offering a formal bridge between information-theoretic and dynamical notions of collective and joint agency.



\section{Formalizing individual and joint agency in flocking behavior through Markov blankets}

We develop a simulation of flocking behavior, in which we consider an ensemble of 100 active inference agents ("birds"), each endowed with identical internal models and each minimizing its local variational free energy. To infer its heading direction, each bird uses observations about the heading directions of its neighbors. This simple mechanism promotes the self-organization of the birds into a collective flocking behavior. It is analogous to models in statistical physics \citep{wu1982potts} and flocking simulations in computer graphics \citep{reynolds1987flocks}, but it is based on local inference rather than on predefined rules (see Section~\ref{sec:methods} for a formal specification of the active inference agents).

\begin{figure}
    \centering
    \includegraphics[width=\linewidth]{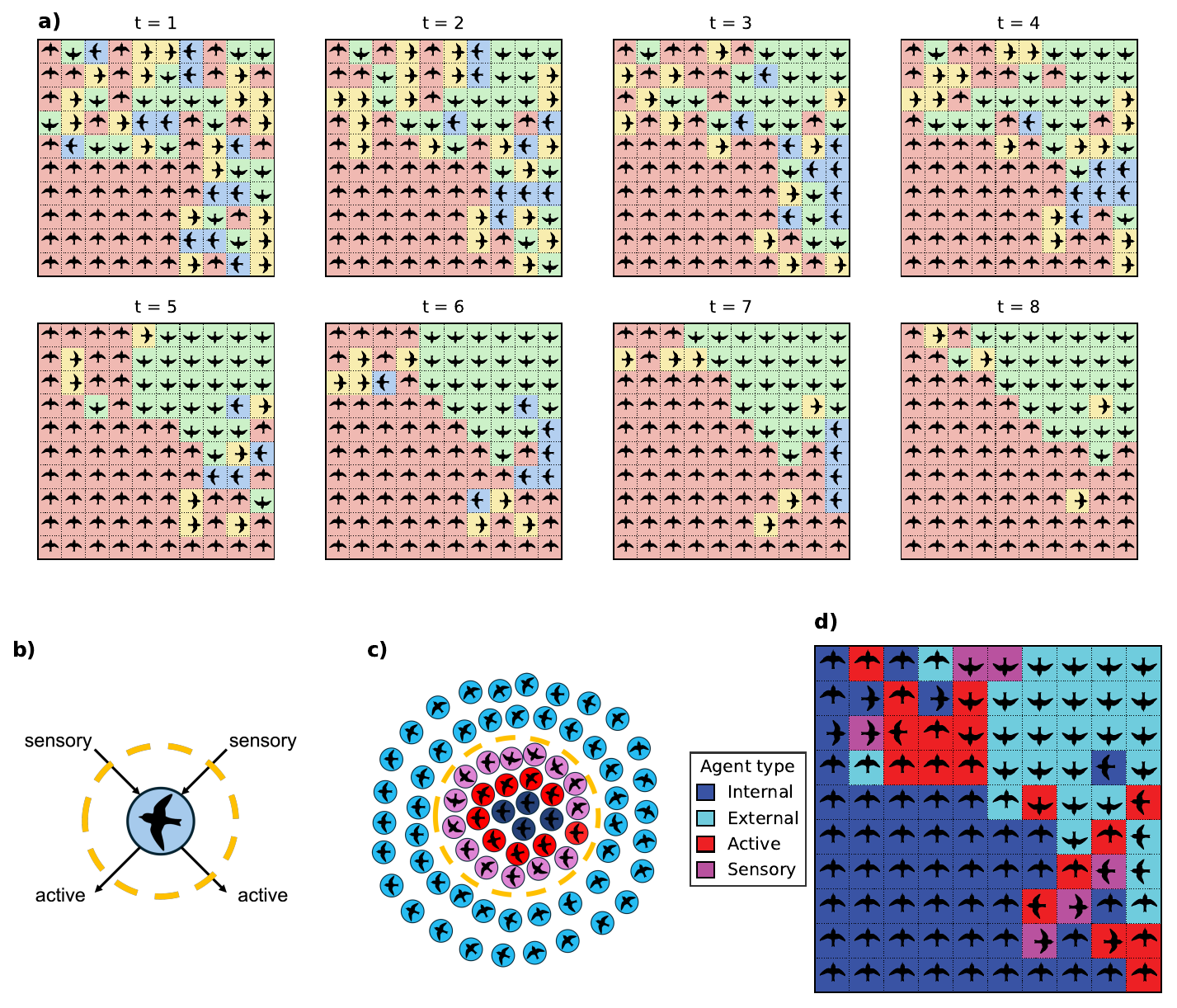}
    \caption{\textbf{Formalizing individual and joint agency in flocking behavior through Markov blankets} (a) Example simulation of flocking behavior among 100 birds over 8 time steps, showing the gradual alignment of headings. (b) Markov blankets, individual and joint agency. Schematic representation of an individual agent, whose internal states are separated from the external environment by a Markov blanket (dashed line). The blanket mediates the exchange of information through sensory $s$ and active $a$ states, defining the agent’s boundary for perception and action. (c) Illustration of a group of interacting agents whose collective dynamics are enclosed within a higher-level Markov blanket (dashed line). Through reciprocal coupling and shared information flow, the ensemble functions as a single, higher-order agent ('flock'), exemplifying the emergence of joint or shared agency from multiple interacting components. Colors illustrate the statistical roles of individual birds relative to the flock Markov Blanket: internal (blue), active (red), and sensory (magenta). External states are depicted as cyan birds. Note that all the birds functioning as internal states are oriented in the same direction, whereas this coordination is not required for birds functioning as sensory and active states. (d) The emergent Markov blanket around the flock during time steps 1--6 of the simulation in (a). The figure illustrates an average over the first 6 timesteps. 
    }
    \label{fig:flocking}
\end{figure}


Figure~\ref{fig:flocking}A illustrates 8 consecutive time steps of an example simulation of flocking behavior among 100 birds, with the colors and orientations of the inset (bird) images indicating each bird's current heading direction. The simulation shows the gradual alignment of the birds' headings over time.

To assess whether this alignment can be formalized as a transition toward joint agency---namely, from individual birds to a \emph{flock}---we turn to the notion of (nested) \emph{Markov blankets} \citep{kirchhoff2018markov,parr2020markov,friston2023path}. A Markov blanket defines a statistical boundary around an agent, delineating the interface between its internal states and the external environment---and thereby its domain of perception and action \citep{pearl2014probabilistic} (Figure~\ref{fig:flocking}B). It is based on the notion of \emph{statistical independence}: internal and external states are conditionally independent given a set of \emph{blanket states}, typically partitioned into sensory and active states. Sensory states mediate the flow of information from the environment to the internal states (observations), while active states mediate the influence of internal dynamics on the environment (actions). This statistical separation enables an agent to maintain its identity and autonomy by inferring hidden causes in the environment from its sensory inputs, and by acting to minimize the discrepancy between predicted and observed states.

Crucially, the Markov blanket formalism allows for \emph{hierarchical nesting} of agents and collectives. When multiple agents become sufficiently coupled through reciprocal sensory and active exchanges, their collective dynamics can give rise to a higher-level Markov blanket encompassing the group as a whole (Figure~\ref{fig:flocking}C). In this configuration, some of the individual agents' states function as the internal, sensory, and active components of a superordinate blanket, thereby defining a new, emergent agent. Thus, while each agent retains its individual agency, the group simultaneously acquires a higher-level, joint agency. Such nesting provides a principled, statistical account of how joint agency can arise naturally from the interactions of individual agents: as informational boundaries reorganize, new levels of autonomy and coordinated inference emerge \citep{friston2013life,heins2024collective}.

Figure~\ref{fig:flocking}D shows the Markov blanket around a collective of birds---which we henceforth call a \emph{flock}---that emerges over time steps 1--6 of the simulation. The colors illustrate that individual birds play distinct roles in the flock, serving as internal (blue), active (red), and sensory (magenta) states, whereas the remaining birds (cyan) constitute external states outside the Markov blanket. This visualization demonstrates that, during the simulation, a statistical separation emerges between birds that are part of the flock and those that are not, thereby providing a formal characterization of joint agency in the flock. Our simulations further demonstrate that the flock preserves certain macroscopic characteristics---such as its overall direction and approximate shape---even as its precise boundaries fluctuate over time. 

Summing up, we have shown that the notion of Markov blanket provides a principled way to formalize the emergence of a collective agent (the flock) above and beyond the individual agents (the birds). The flock can thus be ascribed a form of (joint) agency, grounded in the statistical separation (or autonomy) of its internal states from external states, and in the mediating roles of sensory and active states that couple the two. At the same time, the joint agency of the flock does not diminish the individual agency of its constituent birds, which continue to infer and minimize their own free energy autonomously. These two levels of autonomy, therefore, coexist and are hierarchically nested within one another.


\section{Sensing and escaping predators in the flock}

Having defined the notion of joint agency in a flock, we now assess to what extent the flock can ``sense'' and ``react to'' external perturbations. To this end, we extend the flocking simulation by introducing a ``predator'' that appears at two distinct random positions at time steps 5 and 35, and disappears at the following time step. The predator destabilizes the flock: all birds that sense the predator enter a ``stress'' state that compels them to escape by moving in random directions. Furthermore, this stress state propagates to neighboring birds, causing them also to move randomly and triggering a fast cascade that destabilizes the entire flock.  

This simulation allows us to compare the effects of the predator at two stages: an early stage (step 5), before the birds have self-organized into a large flock, and a later stage (step 35), when a large, cohesive flock has already formed.  

\begin{figure}
    \centering
    \includegraphics[width=\linewidth]{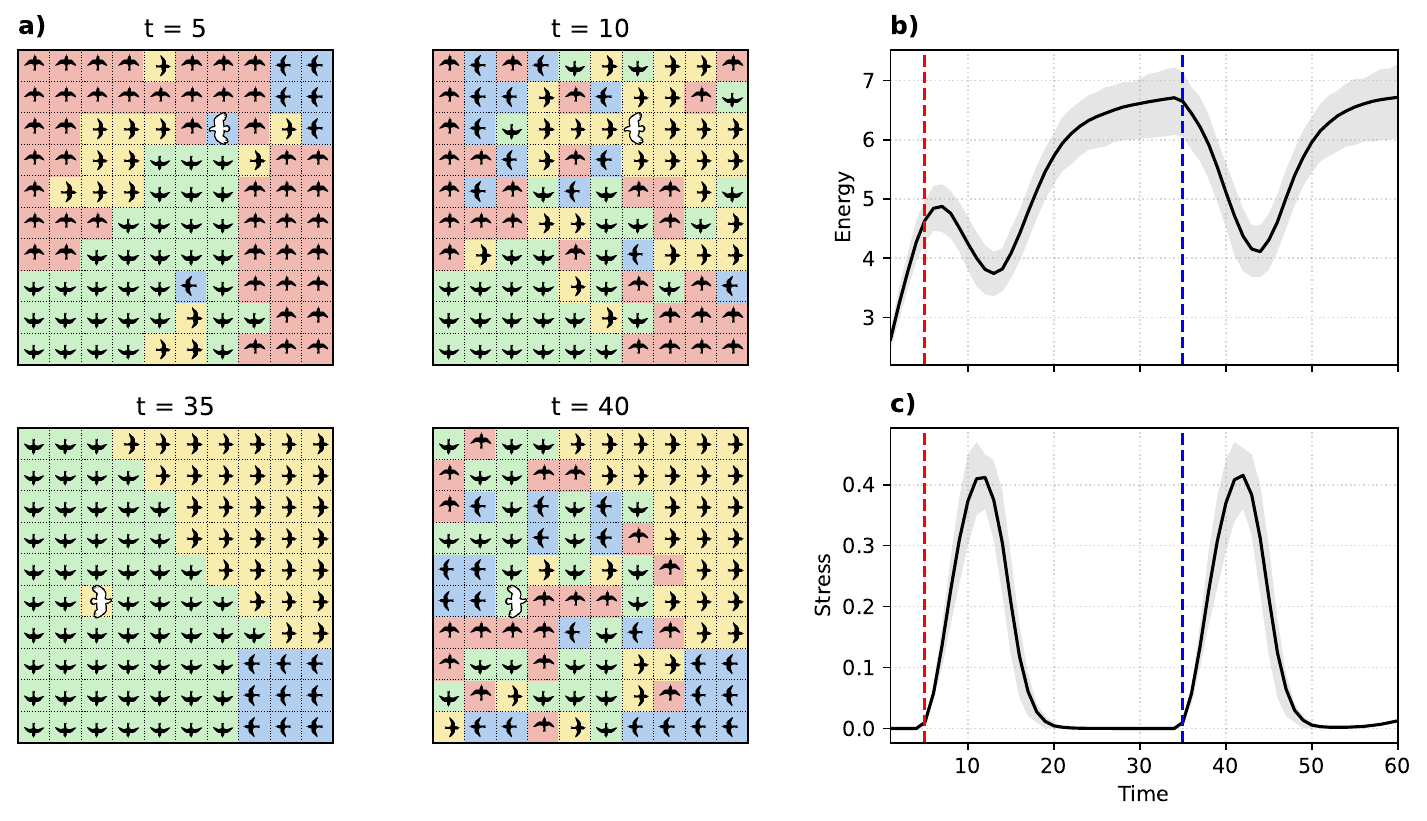}
    \caption{\textbf{Sensing and escaping predators in the flock.} (a) Example of four configurations from a single simulation run. A predator (indicated by a white bird) arrives at time steps $t=5$ and $t=35$, at two distinct, random locations. Configurations are also shown 5 steps following each arrival ($t=10$ and $t=40$). White birds are shown in the plots for illustrative purposes only; they are not part of the simulation except at time steps $t=5$ and $t=35$. (b) System energy, representing the global degree of alignment among agents. The dashed lines show the arrival times of the first (red) and second (blue) predators. The gray band indicates the interquartile range across all simulations. (c) Agents' average stress state over time. The dashed lines show the arrival times of the first (red) and second (blue) predators. The gray band indicates the interquartile range across all simulations. See the main text for further explanation.}
    \label{fig:predator}
\end{figure}

Figure~\ref{fig:predator} shows the results of 500 simulations, each lasting 60 time steps, in which the predator appears at random positions. We excluded from the analysis and the plots 32 simulations where the agents' 'stress' state, induced by the first predator, had not fully returned to baseline before the arrival of the second predator. Figure~\ref{fig:predator}A displays four configurations, before and after the predator appears at time steps 5 and 35, in one representative simulation. It permits visually appreciating that the birds are more organized before the second appearance of the predator and that in both cases, the predator 'destablizes' the system.

Figure~\ref{fig:predator}B shows the energy of the system, defined as in the vector Potts model with four states and Moore neighborhood 
(see Equation~\ref{eq: Potts Hamiltonian} in~\ref{appendix: GM for flocking})
\citep{wu1982potts}, which characterizes the average degree of alignment among the birds (and thus approximates the size or coherence of the flock). 
When the predator appears at an early stage (red dotted line), the energy continues to increase briefly before decreasing, indicating that disorganization occurs with some delay.
In contrast, when the predator appears later (blue dotted line), the energy is initially much higher due to the presence of an already well-formed flock; here, the decrease in energy occurs more rapidly, reflecting a faster predator-induced destabilization.

Figure~\ref{fig:predator}C shows the dynamics of the \textit{grand mean} (i.e., the mean over the simulations of means over birds) of the ``stress'' states (introduced in~\ref{appendix: GM for flocking}) across birds, which—as expected—increase sharply when the predator appears and then gradually decline. This pattern is expected, since the stress propagation is designed to be independent of the agents' current states.

Summing up, this simulation illustrates that predator-induced destabilization occurs faster the second time the predator appears, when a larger flock is present. This occurs despite the propagation mechanism of the ``stress'' state being identical in both cases (Figure~\ref{fig:predator}C). The faster response of the flock may indicate more efficient information propagation within the collective, a hypothesis we investigate in the next analysis.

\section{Synergistic information about predator location in the flock}

In this analysis, we turn to our central question: can we identify a notion of collective knowledge in the flock that extends beyond that of its individual agents?

Specifically, we assess whether the flock as a whole carries information about the predator’s location that individual birds do not possess---and whether this information is greater during the second appearance of the predator (at time step 35, when a larger flock is present) than during the first (at time step 5, when the birds are still close to a random configuration).

To quantify this, we employ \emph{partial information decomposition} (PID) \citep{williams2010nonnegative}, an information-theoretic framework that measures how two source variables (e.g., a pair of birds) together provide information about a target variable (e.g., the location of the predator). Crucially, PID decomposes total mutual information into three components: the \emph{unique information} provided by each source individually, the \emph{redundant information} shared between them, and the \emph{synergistic information} that emerges only when both sources are considered jointly. Synergistic information has been proposed as a foundation for formalizing \emph{emergence}, capturing higher-level informational structure beyond the sum of individual contributions \citep{rosas2020reconciling}.


We compute all four PID atoms across all pairs of birds (source variables) and for each of the two predator locations (target variable), over all 60 time steps and 1000 simulation runs. In each simulation, the predator appears at two distinct random positions. To reduce the dimensionality of the target variable and mitigate estimation bias, we coarse-grained the predator's $10 \times 10$ grid location to a $2 \times 2$ grid. This results in four large, dynamically equivalent quadrants, which remain uniformly sampled by the predator. This coarse-graining ensures that any detected information about the target's location must be dynamically encoded by the agents, rather than stemming from static, pre-existing properties of the environment (e.g., center-vs-edge effects).

\begin{figure}
    \centering
    \includegraphics[width=0.9\linewidth]{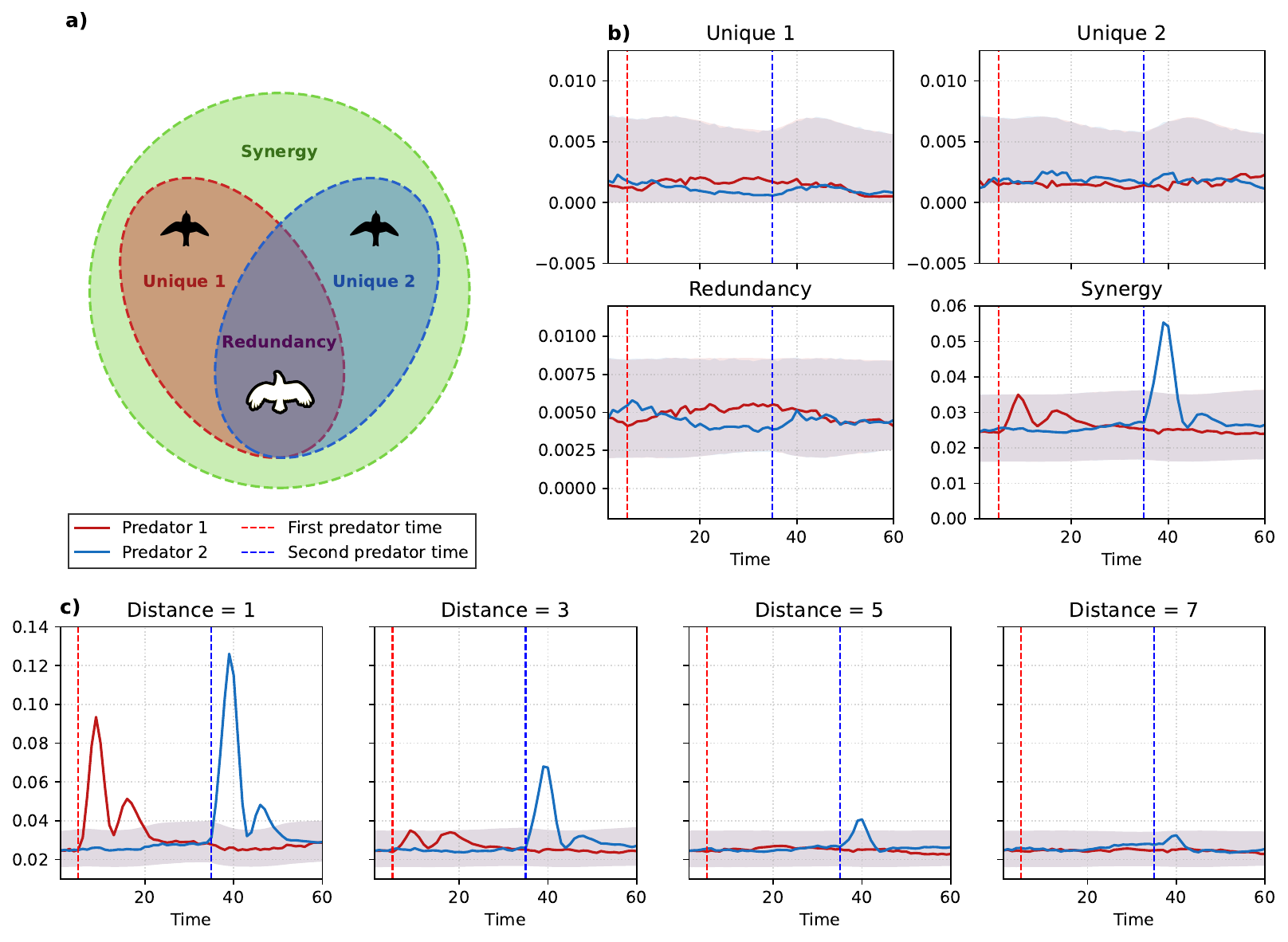}
    \caption{\textbf{Synergistic information about predator location in the flock.}
    (a) Partial Information Decomposition (PID) of the mutual information between two source variables (agent states, $X_1$ and $X_2$, black birds) and a target variable (predator position, in one of four $5 \times 5$ quadrants, $Y$, white bird). The blue and red ellipses delineate the contribution of $X_1$ and $X_2$, respectively, while the green circle shows the combined contribution. The total information is partitioned into the four unique atoms: \emph{redundancy} (the purple overlap) represents information available from both sources, \emph{synergy} represents information available only when both sources are considered together (outside the ellipses), and \emph{unique information} represents information available from only one source.
    (b) Temporal dynamics of the four PID atoms: unique information from the first (top left) and second (top right) agent in each pair, redundancy (bottom left), and synergy (bottom right). Each atom is averaged over 1000 simulations and all 4950 pairs of agents. The red curve shows the case where the target is the position of the first predator, and the blue curve shows the case where the target is the position of the second predator. Vertical dashed lines indicate the arrival times of the predators. The gray band represents the 5th to 95th percentile of the null distribution, obtained by 500 random permutations of the target variable. 
    (c) Temporal dynamics of the synergy as a function of the spatial distance between agent pairs. Each panel shows the dynamics, mediated across the same 1000 simulations, calculated only for agent pairs separated by a specific Chebyshev distance $D = 1, 3, 5, 7$ (from left to right). The figure conventions (red/blue curves for predator targets, vertical dashed lines for arrival times, and gray band for the null distribution derived from target permutations) are identical to those in (b).}
    \label{fig:synergistic}
\end{figure}

Figure~\ref{fig:synergistic} summarizes these results. The core concepts of PID, including unique information, redundancy, and synergy, are formally introduced in~\ref{appendix: IT anlysis} and conceptually illustrated in Figure~\ref{fig:synergistic}A. Figure~\ref{fig:synergistic}B summarizes the temporal dynamics of all four PID atoms, averaged across all bird pairs. Notably, the unique information atoms (top panels) and the redundancy atom (bottom left panel) remain comparable to zero, falling within the gray band of the null distribution at all time steps. In sharp contrast, the synergy atom (bottom right panel) is the only variable that shows significant values. Specifically, we observe a significant peak in synergistic information for the second predator appearance (blue curve). In contrast, the synergy related to the first predator (red curve) shows a similar temporal shape (a peak, decline, and rebound), but it falls within the null distribution band and is therefore not statistically significant. The significant synergy (blue curve) peaks shortly after the predator's appearance and then declines, with a smaller rebound thereafter.

The impact of spatial separation (measured using Chebyshev distance) on the synergy is further detailed in Figure~\ref{fig:synergistic}C. For pairs in close proximity ($\mathrm{Distance}=1$, left panel), significant synergy is detected for both the first (red curve) and second (blue curve) predator appearances. As the distance increases ($\mathrm{Distance}=3$ and $\mathrm{Distance}=5$, middle panels), the synergy for the first predator becomes non-significant, while the synergy for the second predator remains robustly above the null distribution. At the longest distance shown ($\mathrm{Distance}=7$, right panel), synergy becomes non-significant for both events. In all cases, the synergy is higher for the second predator than for the first. 



Summing up, this analysis reveals that the flock exhibits synergistic information about the predator’s location, and that this information is stronger and more spatially distributed during the second predator event, when a larger flock is present. This finding illustrates a form of \emph{collective knowledge} that extends beyond the information available to individual birds, and may help explain the faster collective response observed during the second predator encounter. It also points toward a notion of \emph{causal emergence}, supported by the presence of synergistic information \citep{rosas2020reconciling}.

\section{Discussion}
\label{sec:discussion}

Many of the most fascinating phenomena in biology arise from the self-organization and coordinated behavior of simple agents---cells, neurons, animals, humans, and beyond. Such collective dynamics have been studied across multiple scales, from high-level cognitive constructs such as shared goals and plans, to low-level collective phenomena mediated by reciprocal interactions among simple agents that lack explicit representations of “you,” “we,” or other higher-level cognitive concepts.

In this work, we have explored the notion of \emph{joint agency} in a minimal setting: a population of active inference agents (birds), each minimizing its own free energy while remaining informationally coupled with others. Through simulations of collective self-organization in flocking behavior, we examined three interconnected questions.

In the first analysis, we used the concept of (nested) \emph{Markov blankets} to formalize how agency can extend beyond individual active inference agents to groups of agents---that is, from birds to flocks. We showed that during flocking, the Markov blanket, which formalizes a statistical boundary around an agent, can expand to enclose a set of birds forming a collective unit. Within the flock, subsets of birds play the roles of internal, sensory, and active states of a collective Markov blanket, while others constitute external states. This analysis extends prior work on collective, multi-agent systems---including groups of neurons \citep{palacios2019emergence}, collections of primordial cells \citep{friston2013life}, and developing tissues \citep{friston2015knowing}---by showing that the notion of the Markov blanket can account for situations in which each agent retains its individual autonomy, while the collective simultaneously acquires a higher-level, joint agency. From this perspective, agency is not a fixed attribute of individual systems but an emergent property of coupled dynamics that maintain conditional independence from their surroundings. As interactions among agents strengthen, the boundaries that define “self” can expand or contract, giving rise to transient or stable forms of joint agency. This provides a unified statistical and dynamical framework for understanding how perception, action, and decision-making scale from individuals to collectives---from neurons forming neural assemblies to organisms forming social or ecological systems.

In the second and third analyses, we asked whether being endowed with joint agency---that is, forming a flock---confers additional capabilities or knowledge compared to individual agents. To this end, we introduced a “predator” that appeared at random positions during both early stages of self-organization (when no flock or only a small flock was present) and later stages (when a large flock had formed). The predator destabilized the system by triggering a “stress” state that caused nearby birds to reorient randomly, propagating disruption through the flock.

The simulations revealed that larger flocks reacted more rapidly to the predator, as indicated by a faster onset of destabilization during later stages, even though the rules governing the propagation of “stress” did not depend on the level of organization. Most importantly, our results show that the flock as a whole possesses \emph{synergistic information} about the predator’s location, and that this information increases with flock size and extends farther from the predator. This suggests a formally grounded---albeit implicit---notion of collective knowledge (or shared world model) that goes beyond the information available to individual birds.

Summing up, our simulations illustrate a simple yet general method to characterize a collective agent, such as a flock, as an entity that maintains statistical separation from its environment through a Markov blanket, and exhibits knowledge beyond its components through synergistic information. More broadly, these results illustrate how collective entities can extend the spatial and temporal reach of perception, cognition, and action---or their \emph{cognitive light cone} \citep{levin2019computational}---beyond the boundaries of any single agent.

Future work could extend and increase the realism of the flocking simulation in several ways. Although flocking in nature is a continuous process, here we model it as a discrete one. This choice follows a large body of work, which has shown that flocking dynamics can be accurately captured using discrete formulations based on lattice spin-glass models derived from statistical physics \cite{solon2013revisiting,chatterjee2020flocking}. One advantage of this approach is that, while finite-size effects render numerical studies of continuous models computationally costly and analytical treatments generally complex \cite{ginelli2016physics}, spin-glass formulations are more tractable while preserving key physical features of flocking behavior. In our setting, another key advantage is that the discrete formulation allows us to easily identify regions of aligned elements coexisting with regions of differing or disordered alignment, a distinction that is crucial for our assessment of nested Markov blankets. By contrast, the macroscopic properties of continuous models typically rely on a global transition from disordered to ordered phases, treated as a statistical property of the entire system \cite{bialek2012statistical}. Nonetheless, future work could revisit our findings using continuous-time formulations of multi-agent active inference, as explored in previous studies \citep{friston2013life,friston2015knowing,heins2024collective}. This would allow identification of Markov blankets by examining the spatial clusters that emerge and group agents, offering a complementary perspective to the methodology adopted here.

Future work could also revisit several simplifying assumptions of our simulation to increase biological realism. For instance, our model considers only flocking configurations in which each cell contains exactly one bird. While this assumption appropriately captures high-density regions of the flock—typically composed of interior individuals with uniformly distributed local interactions—it less accurately represents peripheral regions. Moreover, the model assumes a unit distance between neighboring cells. This simplifying assumption is motivated by evidence that interactions governing collective behavior follow a \emph{topological}, rather than metric, framework \cite{ballerini2008interaction}, whereby individuals interact with a fixed number of neighbors rather than maintaining fixed distances. Nevertheless, future work could explore ways to relax these assumptions in order to model more realistic flocking scenarios and to address novel questions, such as how sparseness and inter-individual distance influence the formation and dynamics of Markov blankets and synergistic information about predator location.

Future studies could also address the important conceptual distinction between the implicit form of collective knowledge revealed by synergistic information and the explicit forms typically studied in cognitive science. In cognitive and social sciences, collective knowledge is often assumed to rely on explicit internal models of others’ mental states or shared epistemic representations (e.g., “common ground” or “what we all know that we know”). Similarly, our simulations demonstrate collective sensing and action that do not depend on such explicit representations but instead arise from simple interaction rules. In human societies, collective action is often defined by intentionality---individuals act \emph{with the goal} of producing collective outcomes. By contrast, in simpler biological systems, such as morphogenesis or pattern formation, intentionality can be replaced by teleological organization: systems are structured in such a way to produce collective outcomes even without explicit goals.

An open question, then, is how to bridge our minimal notion of joint agency with more sophisticated accounts used in cognitive science. A natural starting point is the \emph{good regulator theorem} \citep{conant1970every}, which states that any effective controller must embody a model of the system it regulates. In the multi-agent flock we describe, no single bird possesses an explicit generative model of the flock’s collective behavior or its interaction with the environment. Instead, the flock’s generative model is implicitly distributed across the birds and their interaction patterns. In this sense, one can interpret this as the flock not \emph{having} a model of collective action but rather \emph{being}—or dynamically \emph{becoming}—such a model through the structured couplings among its constituent agents. More complex agents, by contrast, possess explicit internal models of collective behavior, with shared representations, goals, and intentions. Understanding how living systems evolve from \emph{being} to \emph{having} a model may shed light on the developmental and evolutionary origins of collective cognition \citep{pezzulo2022evolution}.

Another possible line of work could explore potential cross-fertilizations between the simple simulations presented here and recent efforts to understand more complex forms of collective cognition from an inferential perspective. For example, \emph{Collective Predictive Coding} investigates how agents form communities and develop shared linguistic symbols through decentralized, independent action decisions \citep{taniguchi2025system,taniguchi2025collective,taniguchi2024collective}. Another perspective, \emph{Thinking Through Other Minds}, emphasizes the emergence of shared understanding and social coordination by modeling how agents represent and infer the mental states of others \citep{veissiere2020thinking}. Yet another approach examines long-lasting societal dynamics within groups of agents, using large language models as generative models \citep{park2023generative}. What distinguishes our models from these approaches is the relative simplicity of the agents’ generative models and their (non-linguistic) forms of communication. However, in principle, the common framework of multi-agent generative modeling could allow the same methodology to be applied across a wide range of collective behaviors. Future work could investigate whether and how the methods used here to characterize the emergence of joint agency, (implicit or explicit) collective knowledge, and shared world models in simple multi-agent systems, such as flocks, can be extended to study more sophisticated forms of collective cognition.

Finally, this perspective invites consideration of agency at even broader spatial and temporal scales, such as in \emph{niche construction} \citep{constant2018variational,pio2025scale}. Extending these ideas across generations and cultural dynamics raises deeper questions \citep{pezzulo2025shared,clark1998extended,Friston2024}: To what extent can the nesting of Markov blankets help us understand interactions between individual and social cognition across scales, the social and cultural dynamics of human societies, the formation of collective agency and extended minds, and the ways in which knowledge—distributed across people, artifacts, and institutions such as books, tools, and the internet—shapes individual cognition? We leave these questions open for future research.

\appendix

\section{Supplementary materials}
\label{sec:methods}

\subsection{Active inference framework}

In this work, we consider birds as Bayesian agents with sensory, active, and internal states that are updated according to the Active Inference principles \citep{Parr2022}.
Active inference (AIF) is a theoretical framework that provides a unified account of perception, action, and learning of both living and artificial systems. It posits that an agent's behavior can be understood as the process of maximizing the evidence for the implicit statistical model of the world it embodies, in response to the sensory information flow, by selecting adaptive sequences of actions.

This process is formally grounded in the free-energy principle, according to which, given a physical phenomenon named \textit{generative process}, minimizing a quantity known as variational free-energy corresponds to optimizing the upper bound on the measure of how much the \textit{generative model} used by the agent to describe the phenomenon diverges from the generative process in its predictions.

Following the AIF framework, we designed a generative model describing a single bird that, through local interactions, determines the emergence of flocking behaviors. 

We adopted a hybrid modeling approach. We borrowed the Potts model \citep{wu1982potts}, a statistical physics model describing spin glasses in which spins are arranged on a \textit{lattice}, a periodic graph, and combined it with the computational model developed by Reynolds to mimic collective behaviors in computer vision applications \citep{reynolds1987flocks}.

The Potts model is a generalization of the Ising model \citep{ising1925} for $q>2$ components. It has been used to address numerous problems in collective behavior, which are often described as instances of lattice statistics. Indeed, replacing continuous symmetry with discrete symmetry allows for a simpler, more tractable understanding of the flocking transition. Recently, Solon and Tailleur argued that flocking can be described as a transition from a disordered phase to a polar-ordered phase in an active Ising model, in which spins both diffuse and align on the lattice, a coarse-grained representation of the space \citep{solon2013revisiting}. \citep{chatterjee2020flocking} studied a square lattice in which active particles have four internal states corresponding to the four directions of motion. A local alignment rule inspired by the ferromagnetic 4-state Potts model, combined with self-propulsion via biased diffusion based on the internal particle states, leads to flocking at high densities and low noise.

In contrast, Reynolds introduced a set of prescriptions to force group coordination among multiple agents to simulate herds and swarms. Essentially, Reynolds' rules require each agent to: 
\begin{enumerate}
    \item Attempt to stay close to nearby agent (\textit{flock centering});
    \item Avoid collisions with nearby agent (\textit{collision avoidance});
    \item Attempt to match velocity with a nearby agent (\textit{velocity matching}).
\end{enumerate}
These three rules have a heuristic nature and were conceived to respectively promote cohesion, separation, and alignment of agents known as \textit{boids} -- a compound noun with ``bird'' and the suffix ``oids'' (this latter meaning ``having the likeness of'') -- used to simulate real-life swarms or herds by automated processes. Throughout this section, we will see how these rules are embedded in each agent's generative model and how they influence its internal state, along with observations of neighboring agents' states.

\subsection{Generative model for flocking}\label{appendix: GM for flocking}

Let us consider an ensemble of $N$ birds deployed on a two-dimensional lattice of side $ L = \sqrt {N}$ with coordination number $M = 8$, representing next-nearest-neighbor interactions. 

Each bird is in one of $q$ discrete internal states corresponding to a movement in one of the $q$ lattice directions; just one bird can occupy each site $i$.
In the single bird's generative model occupying the site $i$, the hidden state $z_i$ encodes its propensity to get one of $q=4$ discrete internal states corresponding to cardinal directions.
Each bird is active in the sense that it can flip its internal state and move to a different site. Then, its own control states $u_i$ is an integer in $[0,q-1]$.

The energy of a system represented in this way coincides with the Hamiltonian of a Potts model on a lattice spin glass, defined as
\begin{equation}\label{eq: Potts Hamiltonian}
\mathcal{H} = J_P \sum_{(i,j)} \delta(z_i,z_j)
\end{equation}
where $\delta(z_i,z_j)$ is the Kronecker delta, which equals one whenever $z_i=z_j$ and zero otherwise, $(i,j)$ are the sets of the $M$ nearest neighbor pairs for the indexes $i$, and $J_P$ is a coupling constant that depends on $q$, and is equal for all the potential configurations.

The observations $\tilde{\sigma}_i$ for each single bird consist of the collection of spin states $\sigma_j$ (with $j\in M$ and $j\neq i$) of its $M$ neighboring agents.
In our setup, each bird acts as both the process that generates observations for other birds and the generative model that infers the cause of those observations. Note that the action of one bird constitutes the observed outcomes for another (at the subsequent time step). Therefore, a bird is in a state with a probability corresponding to its beliefs about the average state of its neighborhood at the previous time step. 

We can give the expression of the full predictive generative model of a single bird at time $t$ meant as a joint distribution $P(\tilde{\sigma}_{t},z_t,u_t)$ over the neighborhood observations $\tilde{\sigma}_{t}$, the current hidden state $z_t$ and the related control state $u_t$. One can factorize this distribution to have a form constituted of tractable expressions:
\begin{equation}\label{eq: flocking generative model}
P(\tilde{\sigma}_{t},z_{t},u_t)=P(\tilde{\sigma}_{t}|z_{t})P(z_t|u_t)P(u_t)
\end{equation}
that includes:
\begin{itemize}
     \item A likelihood $P(\mathbf{\tilde{\sigma}}_t|z_t^i)$ encoding neighborly relations.
    Using the three Reynolds' rules to define likelihood establishes how the agents interact. The whole likelihood can be expressed as the product of the single observations $\sigma^j$ received by a single bird in the site $i$ with internal state $z^i$; hence, the likelihood can be factorized as $P(\mathbf{\tilde{\sigma}}|z^i)=\prod_j^M P(\sigma^j|z^i)$ 
    , where:
    \begin{equation}\label{eq: single matrix A}
     P(\sigma^j|z^i;i,j,\beta)= \frac{\exp \left \{-\beta R(i,j)\right\}}{\sum_{k=1}^{M} \exp\left\{-\beta R(i,k)\right\}}
    \end{equation}
    where $\beta$ is the \textit{temperature} of the softmax function, and $R : (i,j) \in \Lambda \times \Lambda \longrightarrow \mathbb{R}$ is a real function defined over the pairs that the site $i$ composes with the neighboring sites, such that:
    \begin{equation}\label{eq: Reynolds function}
        R(i,j) = 
                \begin{cases}
                    v_m, & \left| \eta_i - \zeta_j \right|= 0\\
                    c_a, & \left| \eta_i - \zeta_j \right |=\pi \wedge \theta=\eta_i\\ 
                    f_c, & \left| \eta_i - \zeta_j \right |=\pi \wedge \theta \neq \eta_i\\ 
                    0, & \left| \eta_i - \zeta_j \right | \in \left \{\frac{\pi}{2},\frac{3}{2}\pi\right\}\\ 
                \end{cases}
    \end{equation}
    where $\theta$ is the angle associated to the two-dimensional unit position vector $(\mathbf{r}_j-\mathbf{r}_i)=(\cos \theta, \sin \theta)$, and where 
    $\eta_i$  and $\zeta_j$ are respectively defined as $\eta_i = 2\pi z_i/q$, $\zeta_j=2\pi\sigma_j/q$, with $z_i=0,\dots,q-1$, and $\sigma_j=0,\dots,q-1$. Each one of the real values $v_m$, $c_a$, and $f_c$ is related to a Reynolds' rule; they correspond respectively to the \textit{velocity matching}, \textit{collision avoidance}, and \textit{flock centering} parameters. 
    For instance, setting $v_m>0$ we induce the bird to match the averaged observed state of its neighborhood; setting appropriate values for $c_a$ and $f_c$ we induce in the bird a tendency to penalize heading directions parallel and opposite to the averaged observed state ($\left| \eta_i - \zeta_j \right |=\pi$) both to avoid collisions when the verses of corresponding directions are convergent ($\theta = \eta_i$), and to force the bird to stay in group when verses are divergent ($\theta_{ij} \neq \eta_i$). In our simulations, we used the following parameter values: $v_m=4$, $c_a=2$, and $f_c=1$.
    
    Finally, $R(i,j)$ is zero when the states of the bird and the neighbor are orthogonal, and this circumstance describes configurations where no particular utility, neither rewarding nor penalizing, is assigned. 
    %
    %
    \item A state transition $P(z_t|u_t)$, such that:
    \begin{equation}\label{eq: state transitions without predator}
    P(z_t|u_{t};\rho)=\frac{\exp\{-\rho \, \delta_{z_t,u_{t}}\}}{\sum_u \exp\{-\rho \, \delta_{z_t,u}\}},
    \end{equation}
    with $z_0 \sim \mathbf{Cat}(1/q,\dots,1/q)$ when $t=0$, that assigns to a bird the flight direction specified by the action $u_t$.\\
    
    \item A posterior distribution over control states defined as: \begin{equation}\label{eq: posterior control state}
    P(u_t;\gamma)=\frac{\exp\left\{-\gamma \, \mathbf{G}(u_t)\right\}}{\sum_u \exp \left\{-\gamma \, \mathbf{G}(u_t) \right\}}
    \end{equation}
    where $\mathbf{G}(u_t)$ is the expected free energy of the policy $u_t$ one-control-state long, 
    and $\gamma \in \mathbb{R}$ is a variable denoted as ``active inference precision'', on-line computed to self-tune the control-state selection process adaptively.
    Starting from $\mathbf{G}$ definition as ELBO (Evidence Lower BOund) of the model evidence, it is possible to lead this expression back to already known distributions by using the position $Q(\sigma|z,u) \triangleq P(\sigma|z)$, where $Q$ denotes the \textit{variational approximation} for the generative model. Through easy algebraic manipulations, we can write $\mathbf{G}$ as:
    \begin{gather}
        \begin{split}
               \mathbf{G}(u) &= \mathbb{E}_{Q(z,\sigma|u)} \left[ \log \frac{Q(z|u)}{P(z,\sigma|u)}\right]\\
               &= \mathbb{E}_{Q(z,\sigma|u)} \left[ \log \frac{Q(z|u)}{P(z|\sigma,u)P(\sigma|u)}\right]\\
               & \approx \mathbb{E}_{Q(z,\sigma|u)} \left[ \log \frac{Q(z|u)}{Q(z|\sigma,u)P(\sigma|u)}\right]\\
               &= \mathbb{E}_{Q(z,\sigma|u)} \left[ \log \frac{Q(\sigma|u)}{Q(\sigma|z,u)P(\sigma|u)}\right]\\
               & = \mathbb{E}_{Q(z|u)P(\sigma|z)} \left[ \log \frac{Q(\sigma|u)}{P(\sigma|z)P(\sigma|u)}\right]\\
        \end{split}
    \end{gather}
    
    In the expression that comes out from the last derivation step, $Q(z|u)$ is the state estimation carried out by the outcome predicted at the previous time step, and $Q(\sigma|u)$ is the current outcome belief. 
    In contrast,
    $P(\sigma|z)$ is the likelihood defined in Equation \eqref{eq: single matrix A}, and $P(\sigma|u)$ is a ``goal prior'', i.e., a preferred outcome that renders the preference of being conservative and observing the same state in the future. 
    
    The probability distribution of the predictive future outcome $\tilde{\sigma}_{t}=\{\sigma_{t}^j\}_{j=0,\dots,q-1}$ can be encoded as $P(\tilde{\sigma}_{t}|u_t)\equiv P(\tilde{\sigma}_{t}|\tilde{\sigma}_{t-1})=\prod_{j=1}^{M}P(\sigma_{t}^j|\sigma_{t-1}^j)$, with:
    \[P(\sigma_{t}^j|\sigma_{t-1}^j;\omega) = \frac{\exp \left\{-\omega \, \delta_{\sigma_{t}^j,\sigma_{t-1}^j}\right\}}{\sum_{k=0}^{q-1} \exp \left\{-\omega \, \delta_{k,\sigma_{t}^j}\right\}}\]
    where $\omega \in \mathbb{R}$ is a precision parameter, and $\delta_{\cdot,\cdot}$ denotes the Kronecker delta. Then, in this model, we have assumed that the goal probabilities follow a Boltzmann distribution conditioned by the state of the neighboring birds.
\end{itemize}


\noindent In our simulations, all the temperature parameters $\beta$, $\rho$, and $\omega$ used in the distributions were set equal to 1.

\subsection{A generative model variant for escaping from predator attacks}\label{subsec: escaping from predator attack}

Flocks escape predators through a combination of individual birds and collective behaviors that enable faster reactions and collective escape maneuvers.

Individual vigilance, dilution and detection \citep{foster1981evidence}, the "fountain effect" \citep{hall1986predator}, and confusion \citep{neill1974experiments} are some of the escape strategies that rely on the benefits of being in a flock, where the probability of survival of any one individual increases when its behavior is configurable within a group response.

The choice of escape strategy depends on multiple factors \citep{hilton1999intraflock}
, such as the predator's approach--- the closer the predator gets, the more likely the flock is to initiate an escape response---, the reaction time variation---the distance between predators and individuals influences their reaction time---, and the flock size, connected to predation risk of any individuals as highlithed also by the selfish herd theory \citep{hamilton1971geometry}.

Simulating a collective escape strategy entails extending the original flocking model: each bird must be able to recognize danger and modify behavior in safety situations. To this end, we introduce a second factor, named ``stress state'', in the hidden state that encodes the presence of danger. Therefore, the hidden state of the extended model becomes $z'^i = z^{i} \otimes z^{*i}$, where $z^i$ coincides with the hidden state of the original model, and $z^{*i}$ is a binary variable that indicates being ($z^{*i}=1$) or not ($z^{*i}=0$) in danger.

Analogously, we account for the observation of stress responses by adding, for each $\sigma^{j}$, corresponding to the states of $j$-th neighbor individual, another factor encoding its stress outcome $\sigma^{*j}$, so that $\sigma'^j = \sigma^j \otimes \sigma^{*j}$, ultimately.

At this point, we need to characterize birds' behavior in dangerous situations. We decided to implement the confusion strategy as an escape strategy. In practice, birds in the flock respond to a predator's attack by fleeing at random.
To do this, we need to modify the likelihood matrix $P(\mathbf{\tilde{\sigma}}|z^{i},z^{*i})$ by extending it to the conditional probabilities of the observations $\tilde{\sigma}$ given the hidden state $z^{*i}$. When $z^{*i}=0$, the extended likelihood matrix is still defined as Equation \eqref{eq: single matrix A}, with $\sigma'^j \equiv \sigma^{j}$, while the stress component $\sigma^{*j}$ is uninformative (all the elements have the same value and every column sums to one). In contrast, when $z^{*i}=1$, the component $P(\sigma'^{j}|z^{i},1)$ is analogous to that in Equation \eqref{eq: single matrix A}, taking care to replace the function $R(i,j)$ introduced in Equation \eqref{eq: Reynolds function} with the function $R'(i,j)$ that satisfies the following conditions:
\begin{equation}
\label{eq: Reynolds function for threatening observation}
        R'(i,j) = 
        \left\{
        \begin{array}{lr}
			c_a, & \left| \eta_i - \zeta_j \right|=\pi \wedge \theta=\eta_i\\
			-R(i,j), & \mathrm{o/w}\\
        \end{array} 
                \right.
\end{equation}.

In a way, $z^{*i}$ is a context variable that shapes the perception of the states of neighboring birds and affects the choice of action. Actually, in the latter case, the influence is mutual. Indeed, similarly to the case without predators with $z_t'^i \equiv z_t^{i}$ (see Equation \eqref{eq: state transitions without predator}), there exists a transition probability that probabilistically attributes a value to the stress state to $z_{t}^{i}$, every time an action $u$ is executed. 

Unlike $z_{t}^{i}$, the transition model of $z_{t}^{*i}$ is conditioned by some parameters that entail how the stress state evolves. We presumed that the stress state could follow a specific dynamics that, from the initial threat, leads to a state of no danger present before the attack. Specifically, suppose to denote as $\bar{t}$ the time at which a generic bird has a predator (or a ``stressed'' bird) in its neighborhood. In that case, we hypothesize that its hidden state ${z}_{\bar{t}}^{*i}$ transitions from 0 to 1, then decays to 0 with a mean lifetime constant $\tau$. After that, the individual remains in a quiet state with $z_t^{*i}=0$ for a refractory period $T_r$, during which the individual does not respond to new danger situations. The following transition model can represent the entire dynamics over time:
\begin{equation}\label{eq: transition model for stress state}
P(z_{t}^{*i}|u_t;\bar{t},\tau,T_r)=
\begin{cases}
\begin{pmatrix} 0 & 0\\ 1 & 1 \end{pmatrix}, & \bar{t}<t \le \bar{t}+\tau\\
& \\
\begin{pmatrix} 1 & 1\\ 0 & 0 \end{pmatrix}, & \bar{t}+\tau<t \le \bar{t}+\tau+T_r\\
& \\
\begin{pmatrix} 1 & 0\\ 0 & 1 \end{pmatrix}, & \mathrm{o/w}\\
\end{cases}
\end{equation}
where the columns and rows of the matrices represent the values of $z_{t-1}^{*i}$ and $z_{t}^{*i}$, respectively. It should be noted that Equation \eqref{eq: transition model for stress state} is independent of the action $u_t$ executed and, instead, depends exclusively on the time. In our simulations, we set $\tau=2$ and $T_r=10$.

\subsection{Spectral Identification of Markov Blankets via Fiedler Vector Analysis}

We adopted a spectral approach to detect Markov blankets that characterize groups of birds flying in the same direction, forming a macro-agent.

By fixing a time window $T_w$, we can represent the whole flock as an undirected graph $F=(V,E)$, where the nodes are the birds and the edges denote the fact that two birds fly in the same direction, at least once in $T_w$. We can assume that $A$ is the weighted adjacency matrix of that graph, where $A_{ij}$ is the number of times the nodes $i$ and $j$ are connected (have equal direction). 

The Markov blanket of a subset of nodes $S \subset V$  is defined as the minimal set $M_b(S)$ such that $S$ is conditionally independent of $V \setminus (S \cup M_b(S))$ given $M_b(S)$.

For community detection applications, we consider the Markov blanket separating two communities $C_1$ and $C_2$ as the subset of boundary nodes that minimizes information flow between the communities while maintaining graph connectivity. To find these boundary nodes, we determine the algebraic connectivity (also known as Fiedler eigenvalue) of the graph by computing the second-smallest eigenvalue of the Laplacian matrix of $F$ defined as $L=D-A$ where $D = \text{diag}(\sum_j A_{ij})$ is the degree matrix. The Fiedler vector $y_2$ \citep{fiedler1973algebraic} is the eigenvector corresponding to the second smallest eigenvalue $\lambda_2$ of $L$, such that $L y_2 = \lambda_2 y_2$.

The Fiedler vector provides a natural embedding of nodes along the principal axis of spectral separation. Nodes are projected onto a one-dimensional space where their positions reflect their structural roles: nodes with $|y_2(i)| \approx 1$ represent core members of the communities; nodes with $|y_2(i)| \approx 0$ constitute the Markov blanket between communities.
We identify Markov blanket nodes through thresholding:
\begin{equation}\label{eq: Mb and algebraic connectivity}
M_b = \{ i \in V : |y_2(i)| \leq \alpha \}
\end{equation}
where the threshold $\alpha$ is determined adaptively based on the distribution of $y_2$ values, typically set to capture nodes within the 10th-20th percentile of absolute Fiedler values. 
The quality of a node is determined by its degree of connection to the communities. For instance, taking as a reference the community $C_1$, a node $i$ is internal if $y_2(i)>\alpha$, external if $y_2(i)< -\alpha$, and either active or sensory if, respectively, its connection degree is higher with $C_1$ or $C_2$.

\subsection{Information-Theoretic Analysis of Agent Interactions}\label{appendix: IT anlysis}

Let \(X_{i}^{k}(t)\) denote the state of agent \(i\) at time \(t\) in simulation \(k\), where each simulation is a sample from the same stochastic process, used subsequently to estimate probabilities. The agent’s state is encoded as an integer between 0 and 3, corresponding to its heading direction. The predator position in simulation \(k\) at time \(t\) is denoted by \(Y^{k}(t)\). This variable is discretized into four spatial quadrants (hence also taking values between 0 and 3) and, depending on the analysis, can refer to either the first or the second predator in the simulation.

For each pair of agents \(i\) and \(j\), we compute the univariate and multivariate mutual information, respectively $I_i(t)$ ($I_j(t)$) and $I_{ij}(t)$, between their states \(X_{i}^{k}(t)\), \(X_{j}^{k}(t)\) and the target variable \(Y^{k}(t)\):

\begin{equation}\label{eq: mutual informations}
\begin{aligned}
I_{ij}(t) &= I\bigl(X_i(t), X_j(t); Y(t)\bigr), \\
I_i(t)   &= I\bigl(X_i(t); Y(t)\bigr), \\
I_j(t)   &= I\bigl(X_j(t); Y(t)\bigr),
\end{aligned}
\end{equation}

All mutual information terms are evaluated at each time step $t$, treating \(X_i(t)\), \(X_j(t)\), and \(Y(t)\) as random variables whose realizations are drawn from the ensemble of simulations \(\{(X_{i}^{k}(t), X_{j}^{k}(t), Y^{k})\}_{k}(t)\).

We then perform a standard partial information decomposition (PID) using 
the following expressions:

\begin{equation}
\begin{aligned}
R_{ij}(t) &= \min \bigl\{I_i(t), I_j(t)\bigr\},\\
U_i(t)   &= I_i(t) - R_{ij}(t),\\
U_j(t)   &= I_j(t) - R_{ij}(t),\\
S_{ij}(t) &= I_{ij}(t) - R_{ij}(t) - U_i(t) - U_j(t),\\
\end{aligned}
\end{equation}
where \(R_{ij}\), \(U_i\), \(U_j\), and \(S_{ij}\) represent the redundant --- using the definition present in \citep{williams2010nonnegative} ---, unique (agent \(i\)), unique (agent \(j\)), and synergistic information components, respectively. These quantities are averaged across all agent pairs at a specified interaction distance; in the absence of a fixed distance, they are averaged over all possible pairs.

To assess the statistical significance of the obtained information components, we estimate null distributions by preserving the source variables \(X_i\) and \(X_j\) while applying a temporal shuffle to \(Y(t)\). This procedure is repeated \(N = 500\) times, and the mutual information and PID atoms are recomputed for each surrogate series. This method generates a null distribution under the hypothesis of no systematic information sharing between agent states and the predator’s position, following the surrogate-data approach commonly used in information-theoretic connectivity analysis \citep{vicente2011transfer}.


\section*{Acknowledgments}

This research received funding from the European Research Council under the Grant Agreement No. 820213 (ThinkAhead), the Italian National Recovery and Resilience Plan (NRRP), M4C2, funded by the European Union – NextGenerationEU (Project IR0000011, CUP B51E22000150006, `EBRAINS-Italy'; Project PE0000013, `FAIR'; Project PE0000006, `MNESYS'), and the Ministry of University and Research, PRIN PNRR P20224FESY and PRIN 20229Z7M8N. The GEFORCE Quadro RTX6000 and Titan GPU cards used for this research were donated by the NVIDIA Corporation. We used a Generative AI model to correct typographical errors and edit language for clarity.
\section*{Authors' Contributions}

All authors contributed to the conceptualization and writing of the manuscript. 

\section*{Competing Interests}

We have no competing interests.

\section*{Data availability}

The source code to reproduce the experiments is available at \url{https://github.com/dommai/flock_knows_what_birds_dont}

\bibliographystyle{plain}
\bibliography{sample,bib}

\end{document}